\documentstyle[epsfig,mathptm]{mn}
\newif\ifAMStwofonts
\AMStwofontstrue                
\DeclareMathVersion{bold}                    
\DeclareMathAlphabet{\mathbfit}{OT1}{cmr}{bx}{it}
\SetMathAlphabet\mathbfit{bold}{OT1}{cmr}{bx}{it}
\DeclareMathAlphabet{\mathbfss}{OT1}{cmss}{bx}{n}
\SetMathAlphabet\mathbfss{bold}{OT1}{cmss}{bx}{n}
\ifAMStwofonts
  \ifCUPmtlplainloaded \else
    \DeclareSymbolFont{UPM}{U}{eur}{m}{n}
    \SetSymbolFont{UPM}{bold}{U}{eur}{b}{n}
    \DeclareSymbolFont{AMSa}{U}{msa}{m}{n}
    \DeclareMathSymbol{\upi}{0}{UPM}{"19}
    \DeclareMathSymbol{\umu}{0}{UPM}{"16}
    \DeclareMathSymbol{\upartial}{0}{UPM}{"40}
    \DeclareMathSymbol{\leqslant}{3}{AMSa}{"36}
    \DeclareMathSymbol{\geqslant}{3}{AMSa}{"3E}
  \fi
\fi

\def\hvect#1{{\hat{\bmath{#1}}}} 
 
\title[The effect of point sources on satellite observations of 
the CMB]
{The effect of point sources on satellite observations of 
the cosmic microwave background}
\author[M.P.Hobson et al.]
{M.P.~Hobson$^1$, R.B.~Barreiro$^{2,3}$, L.~Toffolatti$^{4,5}$, 
A.N.~Lasenby$^1$, J.L.~Sanz$^2$, \newauthor
A.W.~Jones$^1$ and F.R.~Bouchet$^6$\\
$^1$ Astrophysics Group, Cavendish Laboratory, 
Madingley Road, Cambridge CB3 0HE, UK\\
$^2$ Instituto de F\'\i sica de Cantabria (CSIC-UC),
Facultad de Ciencias, Av. de Los Castros s/n, Santander 39005, 
SPAIN\\ 
$^3$ Departamento de F\'\i sica Moderna, Universidad de Cantabria,
Facultad de Ciencias, Av. de Los Castros s/n, Santander 39005, SPAIN\\ 
$^4$ Osservatorio Astronomico, Vicolo dell'Osservatorio 5, 35122 
Padova, ITALY\\
$^5$ Departamento de F\'\i sica, Universidad de Oviedo, c.le Calvo 
Sotelo s/n, 33007 Oviedo, SPAIN\\
$^6$ Institut d'Astrophysique de Paris, 98 bis Boulevard Arago, 
75014 Paris, FRANCE}
\date{Accepted ???. Received ???; in original form \today}
\pagerange{\pageref{firstpage}--\pageref{lastpage}}
\pubyear{1998} 

\begin{document}
\maketitle
\label{firstpage}

\begin{abstract}
We study the effect of extragalactic point sources on satellite
observations of the cosmic microwave background (CMB).  In order to
separate the contributions due to different foreground components, a
maximum-entropy method is applied to simulated observations by the
Planck Surveyor satellite.  In addition to point sources, the
simulations include emission from the CMB and the kinetic and thermal
Sunyaev-Zel'dovich (SZ) effects from galaxy clusters, as well as
Galactic dust, free-free and synchrotron emission.  We find that the
main input components are faithfully recovered and, in particular,
that the quality of the CMB reconstruction is only slightly reduced
by the presence of point sources. In addition, we find that it is
possible to recover accurate point source catalogues at each of the
Planck Surveyor observing frequencies.
\end{abstract}

\begin{keywords}
methods: data analysis -- techniques: image processing -- 
cosmic microwave background.
\end{keywords} 
 
\section{Introduction}

A new generation of cosmic microwave background (CMB)
satellite missions are currently in the final stages of design.  The
NASA MAP satellite is expected to be launched by 2000, followed by the
ESA Planck Surveyor in 2007 (Bersanelli et al. 1996). 
Both missions will provide detailed
all-sky maps of the CMB anisotropies, leading to definitive
measurements of the CMB power spectrum. This should allow tight
constraints to be placed on fundamental cosmological parameters and
distinguish between competing theories of structure formation in the
early Universe such as inflation and topological defects.

The maps produced by these satellites will, however, contain
contributions from various foreground components, most notably
Galactic dust, free-free and synchrotron emission as well as the
kinetic and thermal SZ effects from galaxy clusters. In addition,
significant contamination from extragalactic point sources is also
expected. It is therefore clear that in order to obtain maps of the
CMB anisotropies alone, it is necessary to separate the emission due
to these various components.

In a previous paper, Hobson et al. (1998) (hereafter Paper I) use a
non-linear maximum-entropy method (MEM) to separate the
emission due to the different foreground components from simulated
Planck Surveyor observations of a $10\times 10$ deg$^2$ field 
(see Bouchet et al. 1997; Gispert \& Bouchet 1997).
It was found that faithful reconstructions may be produced
not only of the CMB anisotropies but also of the Galactic components and
the kinetic and thermal SZ
effects from massive clusters.  The
analysis presented in Paper I does not, however, include the effect of
extragalactic point sources, since no reliable simulations were
available at that time.

In this paper, we apply a generalised version of the MEM separation
algorithm to simulated Planck Surveyor
observations that also include a population of point sources.
This allows us to assess the
effect of point sources on the accuracy with which the CMB
anisotropies and foreground
components can be recovered.  The simulated maps of the point sources
were created for each Planck Surveyor observing frequency using
the updated number counts in each frequency band predicted by
Toffolatti et al. (1998).  These predictions are based on the Danese
et al. (1987) model for the evolution of radio selected sources
(adopting an average spectral index $\alpha=0$ for compact
flat-spectrum sources up
to $\simeq$200 GHz and a break to $\alpha=0.7$ at higher frequencies)
and on the model C of Franceschini et al. (1994) for the evolution of
far-IR selected sources, updated by Burigana
et al. (1997) to account for the isotropic sub-mm component estimated
by Puget et al. (1996).  The point source map at each frequency
was produced by Poisson distributing all the sources in the 
$10^{-5}$ Jy $< S(\nu)< 10$ Jy flux range.  Unless sources are identified and
subtracted down to faint flux limits, the contribution of clustered
sources to the CMB anisotropies is generally found to be small in
comparison with the Poisson term.  As discussed by Toffolatti et
al. (1998), far-IR selected sources should dominate the number counts
of bright objects at frequencies $\nu>$ 300 GHz, whereas radio
selected sources should dominate at lower frequencies.  The two
underlying parent populations are largely different and uncorrelated,
consisting of normal spirals and starburst galaxies in the far-IR and
mostly compact AGN, blazars and quasars in the radio.

\section{The separation algorithm}
\label{sepalg}

Our aim is to reconstruct the CMB anisotropies and the foreground
components in the presence of both instrumental noise and contamination
due to extragalactic point sources. A straightforward initial step
might be to subtract the point-source contribution from those pixels
that are obviously contaminated. The procedure can, however, lead to
difficulties, since a particular contaminated pixel can usually only
be identified in a few of the frequency channels.  Therefore, since
the spectral behaviour of point sources is generally rather
complicated, any interpolation between frequencies must be performed
with caution.  An alternative Fourier-filter approach to removing
point sources from CMB maps has been proposed by 
Tegmark \& de Oliveira-Costa (1998).

In our approach, we do not remove any contaminated
pixels from the simulated data. Instead, we consider the point sources
as an additional contribution to the `noise' and allow the generalised
version of the MEM algorithm to reconstruct
the CMB and foreground emission in the presence of the point sources.
The MEM technique is discussed in detail in Paper I and so here we will
outline the basic points and instead concentrate on how to address
the effect of point sources within the existing formalism. 

\begin{figure*}
\centerline{\epsfig{
file=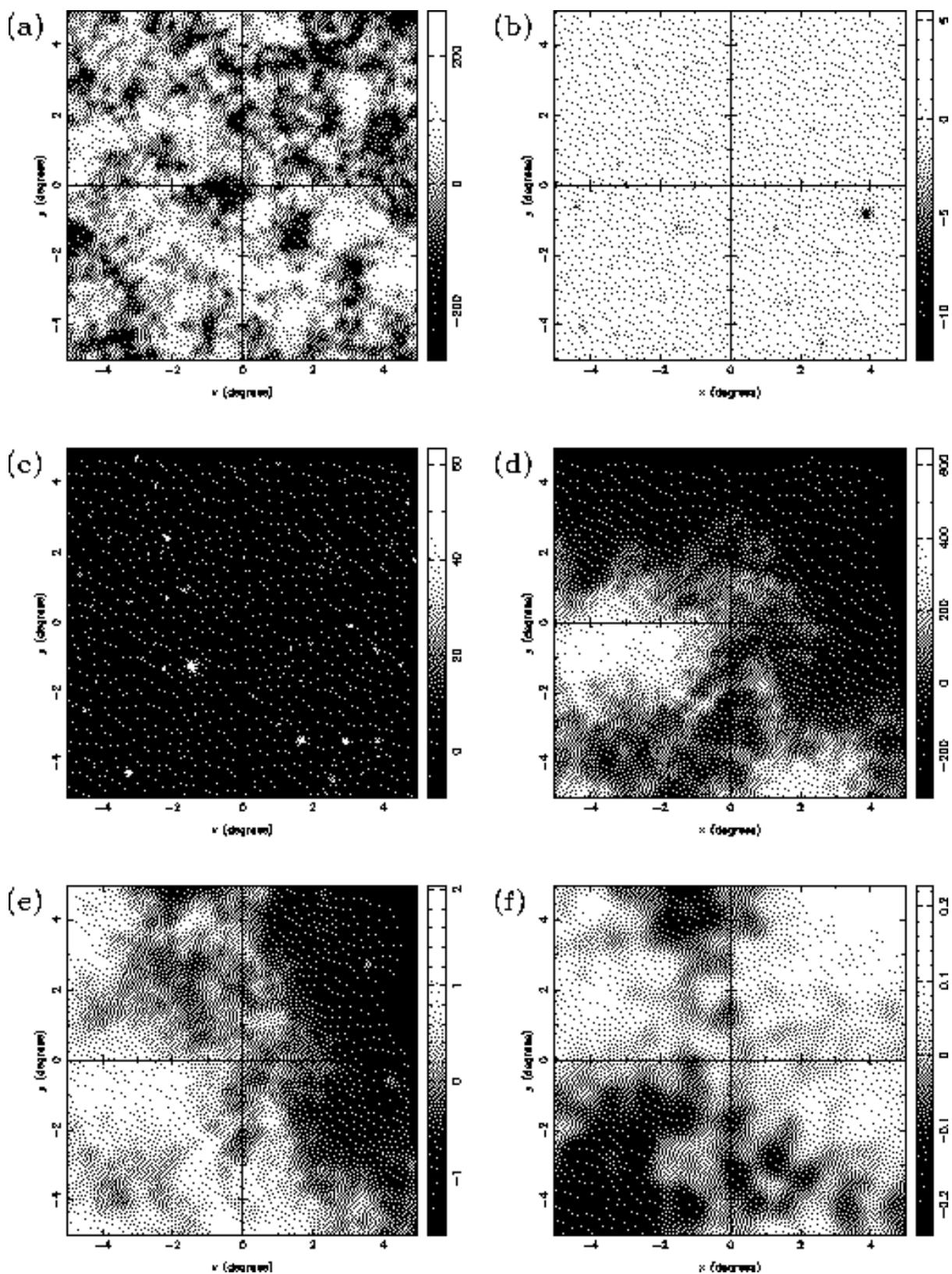,width=16cm}}
\caption{The $10\times 10$ deg$^2$ realizations of the six input
components used to make simulated Planck Surveyor observations:
(a) primary CMB fluctuations; (b) kinetic
SZ effect; (c) thermal SZ effect; (d) Galactic dust; (e) Galactic
free-free; (f) Galactic synchrotron emission. 
Each component is plotted at 300 GHz
and has been convolved with a Gaussian beam of FWHM equal to 4.5
arcmin, the maximum angular resolution proposed for the Planck
Surveyor. The map units are equivalent thermodynamic temperature in
$\mu$K.}
\label{inputs}
\end{figure*}

At any given frequency $\nu$, the total rms temperature fluctuation on
the sky in a direction $\hvect{x}$ is given by the superposition of
the physical components.  In Fig.~\ref{inputs} we show the 
$10\times 10$ deg$^2$ simulated
input maps for all the different physical components (except the point
sources) at a reference frequency $\nu_0 = 300$ GHz; each map 
consists of $400\times 400$ pixels and has been
convolved with a Gaussian beam with a FWHM of $4.5$ arcmin, which is
the highest angular resolution of the Planck Surveyor satellite.
The cell size for each map is 1.5 arcmin.

\begin{figure*}
\centerline{
 \epsfig{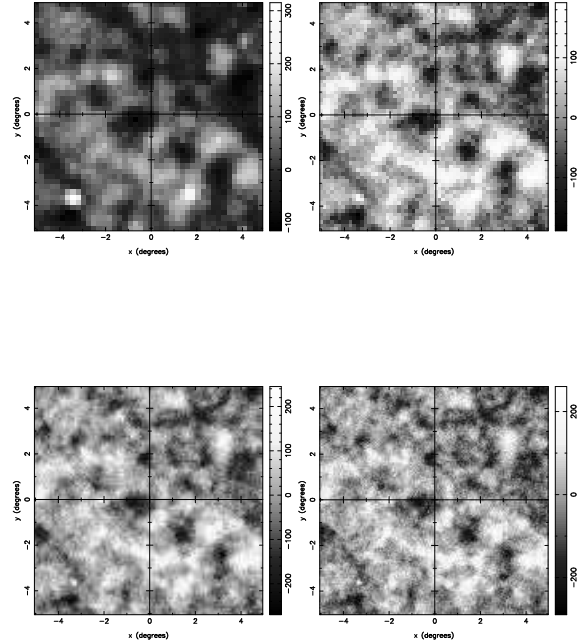}
}
\caption{The $10\times 10$ deg$^2$ maps observed at each of the four
observing frequencies of the Planck Surveyor Low Frequency Instrument
(LFI): (a) 30 GHz, (b) 44 GHz, (c) 70 GHz, (d) 100 GHz.  
At each frequency we assume a Gaussian beam with the
appropriate FWHM and a sampling rate of FWHM/2.4. 
Isotropic noise with the relevant rms has been added
to each map. The map units are equivalent thermodynamic temperature in
$\mu$K.}
\label{data1}
\end{figure*}

\begin{figure*}
\centerline{\epsfig{
file=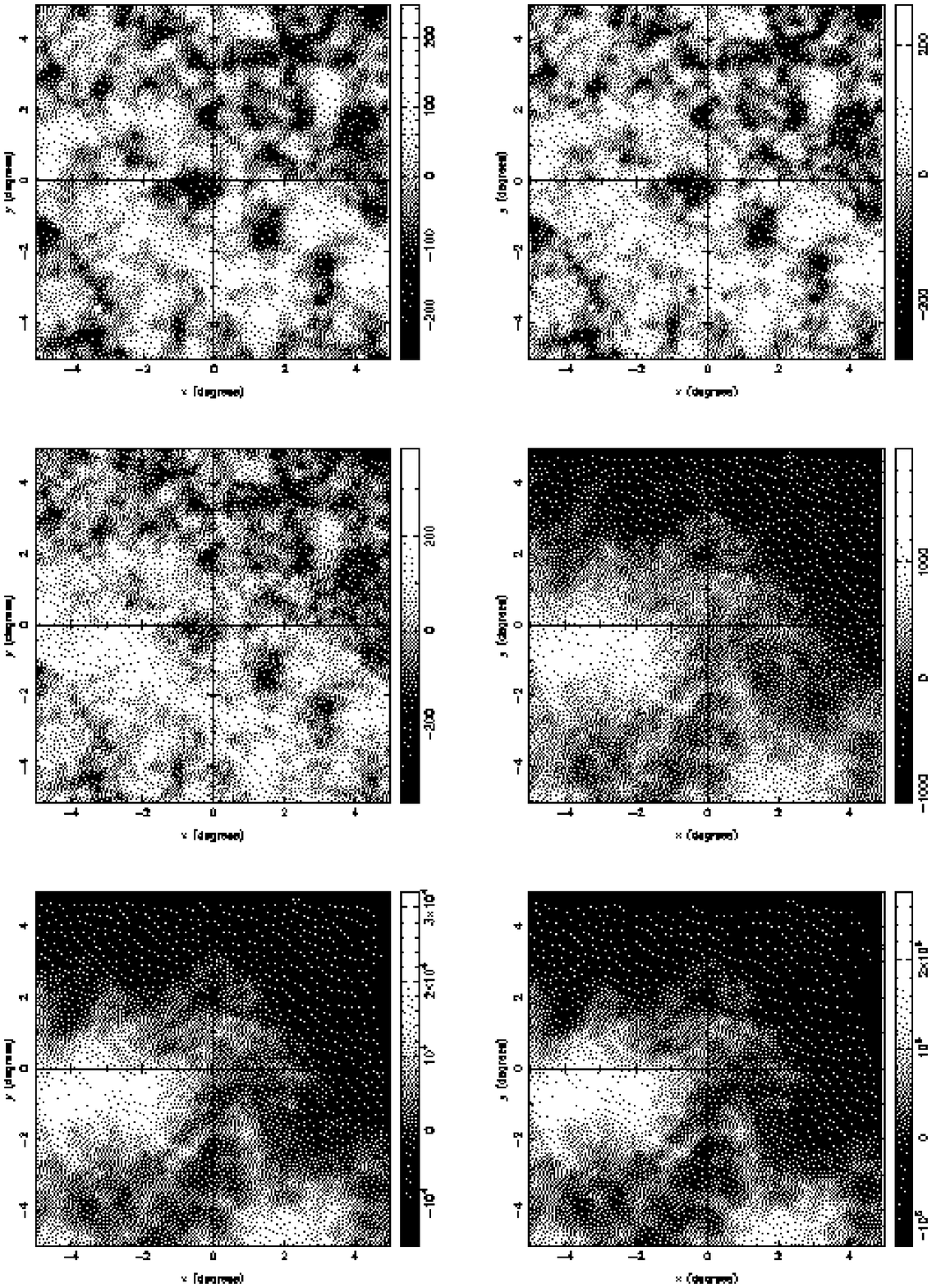,width=16cm}}
\caption{As in Fig.~\ref{data1}, but for the six observing frequencies
of the Planck Surveyor High Frequency Instrument (HFI): (a) 100 GHz,
(b) 143 GHz, (c) 217 GHz, (d) 353 GHz, (e) 545 GHz, (f) 857 GHz.}
\label{data2}
\end{figure*}

If we observe the microwave sky in a given direction $\hvect{x}$ at
$n_f$ different frequencies, we obtain an $n_f$-component data vector
that contains the observed temperature fluctuations in this direction
at each observing frequency plus instrumental noise. 
The $\nu$th component of the data 
vector in the direction $\hvect{x}$ may be written as (see Paper I)
\begin{equation}
d_{\nu}(\hvect{x})=\sum_{j=1}^{N_p} P_\nu(|\hvect{x}-\hvect{x}_j|)
\sum_{p=1}^{n_c} F_{\nu p}\,s_p(\hvect{x}_j) + \eta_\nu(\hvect{x}) 
+ \epsilon_\nu(\hvect{x}),
\label{datadef}
\end{equation}
where $N_p$ denotes the number of pixels in each of the simulated
input maps shown in Fig.~\ref{inputs}.
In this expression we have chosen to separate the contributions from
the six physical components shown in Fig.~\ref{inputs} and the point
sources. The former are collected together in a signal vector with
$n_c=6$ components, such that $s_p(\hvect{x})$ is the signal from the
$p$th physical component at the reference frequency $\nu_0 = 300$
GHz. The corresponding total emission at the observing frequency $\nu$
is then obtained by multiplying the signal vector by the $n_f \times
n_c$ frequency response matrix $F_{\nu p}$ that includes the spectral
behaviour of the considered components as well as the transmission of
the $\nu$th frequency channel. This contribution is then convolved
with the beam profile $P_\nu(\hvect{x})$ of the relevant channel.
Since the spectral dependence of the point sources is very
complicated, we cannot factorize their contribution in this way and
so they are added into the formalism an extra `noise' term. Thus
$\eta_\nu$ is the contribution from point sources as observed by the
Planck Surveyor satellite at the frequency $\nu$. Finally, $\epsilon_\nu$ is
the expected level of instrumental noise in the $\nu$th frequency
channel and is assumed to be Gaussian and isotropic.
Figs~\ref{data1} and \ref{data2} show 
the simulated maps at each observing frequency for the LFI and
HFI respectively. The simulated data are identical to those presented
in Paper I, but also include the point source simulations of Toffolatti
et al. (1998).

The assumption of a spatially-invariant beam
profile in (\ref{datadef}) allows us to perform the reconstruction
more effectively by working in Fourier space, since we may consider
each $\bmath{k}$-mode independently (see Paper I).  
Thus, in matrix notation, at each mode we have
\begin{equation}
\mathbfss{d} 
= \mathbfss{R} \mathbfss{s}+ \boldeta +\bepsilon
= \mathbfss{R} \mathbfss{s}+ \bzeta,
\label{dataft2}
\end{equation}
where $\mathbfss{d}$, $\boldeta$ and $\bepsilon$ are column vectors
each containing $n_f$ complex components and $\mathbfss{s}$ is a
column vector containing $n_c$ complex components. In the second
equality we have combined the instrumental noise vector $\bepsilon$
and the point-source contribution $\boldeta$ into a single vector
`noise' vector $\bzeta$. The response matrix $\mathbfss{R}$ and has
dimensions $n_f\times n_c$ and its elements are given by $R_{\nu
p}(\bmath{k}) = \widetilde{P}_\nu(\bmath{k})F_{\nu p}$.

As discussed in Paper I, the MEM formalism should not itself induce
correlations between elements of the reconstructed signal
vector. However, the elements of the signal vector $\mathbfss{s}$ at
each Fourier mode may well be correlated, this correlation being
described by the $n_c\times n_c$ signal covariance matrix
$\mathbfss{C}$ defined by
\begin{equation}
{\mathbfss C}(\bmath{k}) = \langle 
{\mathbfss s}(\bmath{k})
{\mathbfss s}^\dagger (\bmath{k})
\rangle,
\label{covdef}
\end{equation}
where the dagger denotes the Hermitian conjugate. 
At each Fourier mode, the $p$th diagonal element contains the value of
the ensemble-averaged power spectrum of the $p$th physical component
at the reference frequency $\nu_0$.
Moreover, if prior information is available concerning these
correlations, we would wish to include it in our analysis. 
We therefore introduce the vector of `hidden' variables ${\mathbfss h}$,
related to the signal vector by
\begin{equation}
\mathbfss{s}=\mathbfss{L}\mathbfss{h},
\label{icfdef}
\end{equation}
where the $n_c \times n_c$ lower triangular matrix $\mathbfss{L}$ 
is obtained by performing a Cholesky decomposition
of the signal covariance matrix 
$\mathbfss{C}=\mathbfss{L}\mathbfss{L}^{\rm T}$. 
The reconstruction is then performed entirely in terms of
$\mathbfss{h}$ and the corresponding reconstructed signal vector is 
subsequently found using (\ref{icfdef}).

Using Bayes' theorem,
we choose as our estimator $\hat{\mathbfss h}$ of the hidden vector
to be that which maximises the posterior probability given by
\begin{equation}
\Pr({\mathbfss h}|{\mathbfss d}) 
\propto \Pr({\mathbfss d}|{\mathbfss h})\Pr({\mathbfss h})
\label{bayes}
\end{equation}
where $\Pr({\mathbfss d}|{\mathbfss h})$ is the likelihood of obtaining
the data given a particular hidden vector and 
$\Pr({\mathbfss h})$ is the prior probability that codifies our
expectations about the hidden vector before acquiring any data.

We assume an 
entropic prior probability for the hidden vector $\mathbfss{h}$ of the form
\begin{equation}
\Pr({\mathbfss h}) \propto \exp[\alpha S({\mathbfss h},{\mathbfss m})]
\label{prior}
\end{equation}
where $S({\mathbfss h},{\mathbfss m})$ is the cross entropy of the 
complex vectors $\mathbfss{h}$ and $\mathbfss{m}$, where
${\mathbfss m}$ is a model vector to which ${\mathbfss h}$ defaults in 
absence of data. The form of the cross entropy for complex images 
and the Bayesian method for fixing the regularising parameter $\alpha$
are discussed in Paper I.

Let us now turn our attention to the form of the likelihood function 
$\Pr({\mathbfss d}|{\mathbfss h})$ in Bayes' theorem (\ref{bayes}). 
In the presence of instrumental noise alone, it is
reasonable to assume that the noise contribution is well described by
a Gaussian distribution, but in our case the noise term also
contains a contribution from the point sources, and so we might expect
that a Poisson distribution would be more appropriate for this
component. However, since we are performing the reconstruction in the Fourier
domain, the noise on each $\bmath{k}$-mode will contain contributions
from a wide range of scales. Therefore, provided the point sources are
distributed on the sky in a statistically-homogeneous
manner, we would expect from the central limit theorem that the noise
in the Fourier domain is described, at least approximately, by a
Gaussian distribution. Therefore, we assume that the likelihood
function is given by
\begin{eqnarray}
\Pr({\mathbfss d}|{\mathbfss h}) 
& \propto & \exp \left(-\bzeta^\dag {\mathbfss N}^{-1} \bzeta \right)
\nonumber \\
& \propto & \exp \left[-({\mathbfss d}-{\mathbfss RLh})^\dag 
{\mathbfss N}^{-1} ({\mathbfss d}-{\mathbfss RLh})\right]
\label{likehd}
\end{eqnarray}
where in the last line we have used (\ref{dataft2}). 
The noise covariance matrix ${\mathbfss N}$ has 
dimensions $n_f \times n_f$ and at any given $\bmath{k}$-mode is given
by
\begin{equation}
{\mathbfss N}(\bmath{k}) = \langle\bzeta(\bmath{k})
\bzeta^\dagger (\bmath{k})\rangle.
\end{equation}
Therefore, at a given Fourier mode, the $\nu$th diagonal element of
${\mathbfss N}$ contains the ensembled-averaged power spectrum at that
mode of the instrumental noise plus the point source contribution to
the $\nu$th frequency channel. The off-diagonal terms give the
cross-correlations between different channels; if the noise is
uncorrelated between channels, only the point sources contribute to
the off-diagonal elements.

The argument of
the exponential in the likelihood function (\ref{likehd}) may 
be identified as (minus) the standard $\chi^2$
misfit statistic, so we may write $\Pr({\mathbfss d}|{\mathbfss h})
\propto \exp[-\chi^2({\mathbfss h})]$. Substituting this expression,
together with that for the prior probability given in (\ref{prior}),
into Bayes' theorem, we find that maximising the posterior probability
$\Pr({\mathbfss h}|{\mathbfss d})$ with respect to $\mathbfss{h}$ is
equivalent to minimising the function
\[
\Phi({\mathbfss h})=\chi^2({\mathbfss h}) 
- \alpha S({\mathbfss h},{\mathbfss m}).
\]
This minimisation can be performed using a variable metric minimiser
(Press et al. 1994) and requires only a few minutes of CPU time
on a Sparc Ultra workstation.

\section{Application to simulated observations}

We now apply the MEM analysis outlined above to the simulated Planck
Surveyor data shown in Figs~\ref{data1} and \ref{data2}. 
Clearly, the quality of the reconstructions will
depend on our prior knowledge of the input components. Following Paper I,
we assume that the frequency spectrum behaviour of the components is accurately
known, but we consider two extreme cases relating to our knowledge of
the power spectra of the input components and the confusion noise due
to point sources.

For the six input components shown in Fig.~\ref{inputs}, any prior
power spectrum information may be incorporated into the algorithm
through the signal covariance matrix $\mathbfss{C}$ given in
(\ref{covdef}). Strictly speaking, since we are in fact reconstructing
the vector of hidden variables $\mathbfss{h}$, rather than the signal
vector $\mathbfss{s}$, this information actually resides in the matrix
$\mathbfss{L}$ given in (\ref{icfdef}).  Similarly, any knowledge of
the correlation structure of the instrumental noise and point source
contamination can be included via the noise covariance matrix
${\mathbfss N}$.

\subsection{Reconstructions with full power spectrum information}

As our first case, we assume knowledge of the
azimuthally-averaged power spectra of all six input components in
Fig.~\ref{inputs}, together with the azimuthally-averaged cross power
spectra between components; these contain cross-correlation
information in Fourier space, so that the matrix $\mathbfss{L}$ is
fully specified.  We also assume that the noise covariance matrix
${\mathbfss N}$ is fully specified as follows. In addition to 
the (diagonal) part of the noise matrix due to the instrumental noise, we
calculate an empirical estimate of the contribution (to the diagonal
and off-diagonal elements) of the point sources. This estimate is
calculated directly from 10 independent realizations of the point
source maps at each of the 10 Planck Surveyor frequency channels.

\begin{figure*}
\centerline{\epsfig{
file=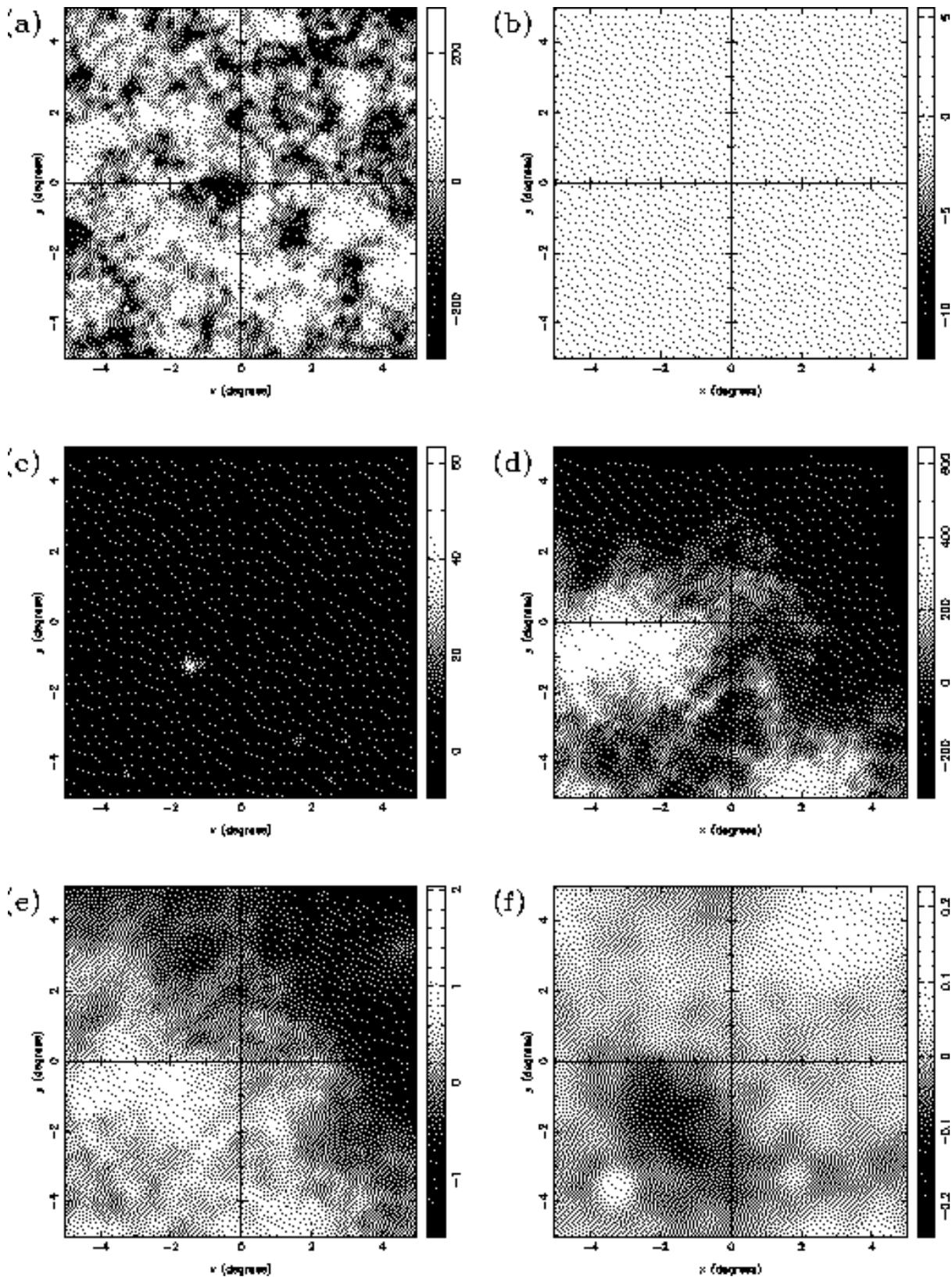,width=16cm}}
\caption{MEM reconstruction of the $10\times 10$ deg$^2$ maps of the
six input components shown in Fig.~\ref{inputs}, using full power spectrum
information (see text). The components are: (a) primary CMB
fluctuations; (b) kinetic SZ effect; (c) thermal SZ effect; (d)
Galactic dust; (e) Galactic free-free; (f) Galactic synchrotron
emission.  Each component is plotted at 300 GHz and has been convolved
with a Gaussian beam of FWHM equal to 4.5 arcmin. The map units are
equivalent thermodynamic temperature in $\mu$K.}
\label{maps1}
\end{figure*}

The resulting MEM reconstructions of the physical components are shown
in Fig.~\ref{maps1} and are plotted on the same greyscale as in
Fig.~\ref{inputs} to allow a straightforward comparison with the true
input maps. 
\begin{figure*}
\centerline{\epsfig{
file=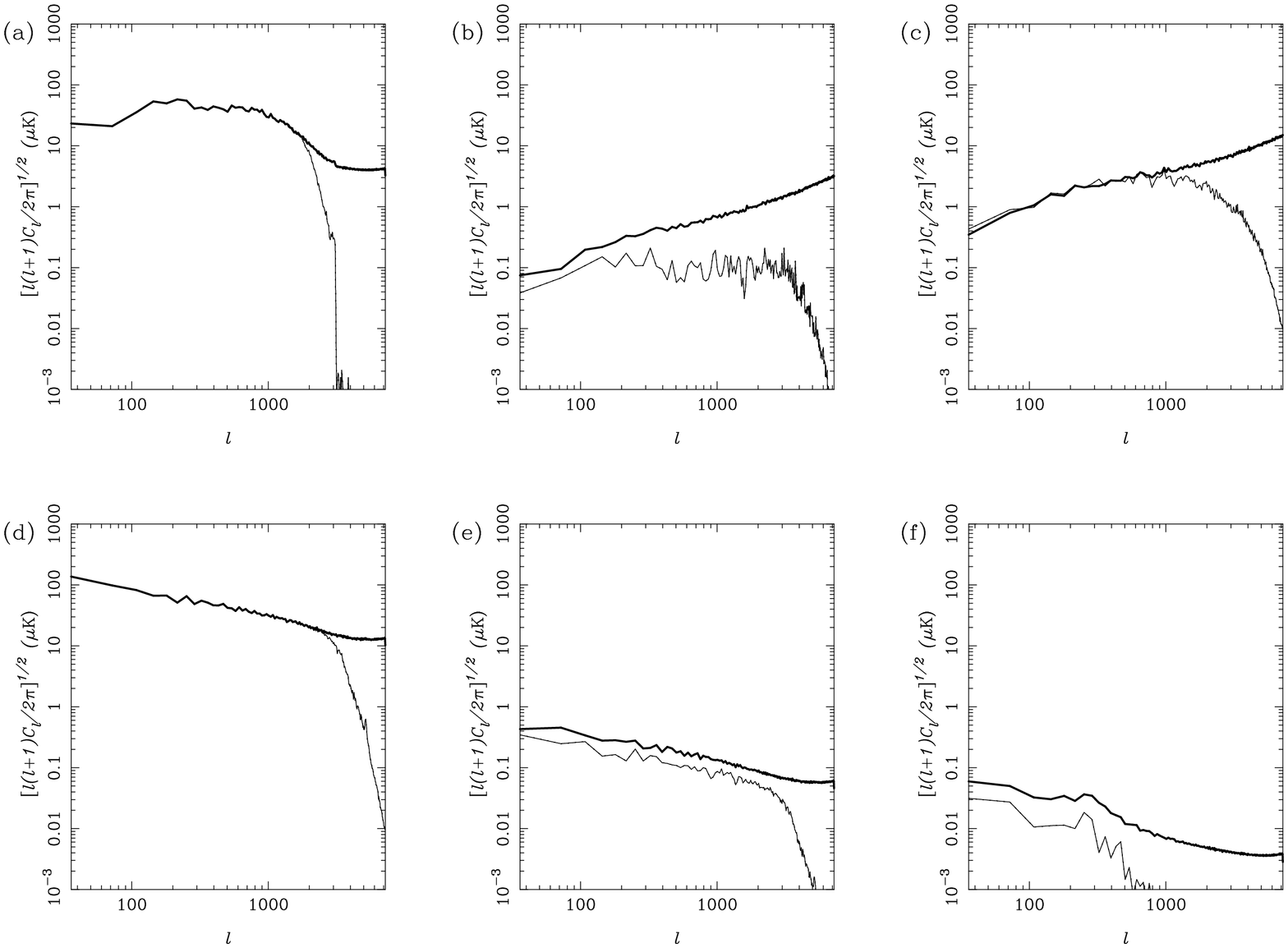,angle=-90,width=14.5cm}}
\caption{The power spectra of the input maps (bold line) compared to
to the power spectra of the maps reconstructed using MEM (thin line). 
In this case the component separation was performed assuming knowledge
of the individual power spectra of the input components.}
\label{ps1}
\end{figure*}
We see that the main input components are faithfully
reconstructed, except for a few contaminated pixels containing the
brightest point sources. Perhaps most notable is the fact that the
majority of contaminating point sources present in the data maps,
shown in Fig.~\ref{data1} and \ref{data2}, are {\em not} present in
the reconstructions. 

In particular, the CMB has been reproduced
extremely accurately and is virtually indistinguishable from the true
input map, apart from a spurious bright point source in the lower
left-hand corner of the reconstruction.  As we might expect the dust
emission is also accurately recovered, and contains none of the
numerous point sources present in the high-frequency data maps shown
in Fig.~\ref{data2}.  The free-free emission, which is highly
correlated with the dust, has also been reconstructed reasonably
accurately and displays no evidence of point source contamination. The
recovery of the synchrotron emission is, however, somewhat poorer.
Although the main features of the true input map are recovered, the
reconstruction contains obvious residual point sources. Perhaps most
impressive is the reconstruction of the thermal SZ effect, since the
MEM algorithm has been successful in reproducing the effect in most of
the bright clusters and has misidentified only a few point sources as
clusters. At the reference frequency of 300 GHz, these misidentified
point sources appear mostly as negative features.
The kinetic SZ effect has also been reconstructed in a
number of clusters having bright thermal SZ effects. In fact, the
quality of the kinetic SZ reconstruction is similar to that obtained
in Paper I.

In order to quantify the effect of point sources on the accuracy of
the reconstructions, in Table~\ref{tab1} we compare the
rms residuals, $e_{\rm rms}$, per
4.5 arcmin FWHM Gaussian beam, of the
reconstructions in Fig.~\ref{maps1}
with those obtained in Paper I in the absence of point
sources. As we might expect, for each component the value of 
$e_{\rm rms}$ is slightly larger when the effects of point sources
are included, but the errors are still reasonable small. In
particular, we note that the rms error on the CMB reconstruction
only increases from 5.9 to 7.5 $\mu$K. 
\begin{table}
\begin{center}
\caption{The rms of the residuals per 4.5 arcmin FWHM Gaussian beam 
(in $\mu$K) of the MEM
reconstructions in the presence and absence of point sources, assuming
full power spectrum information.}
\begin{tabular}{lcc}
\hline
Component & $e_{\rm rms}$ no sources & $e_{\rm rms}$ with sources \\
\hline
CMB          & 5.90 & 7.50 \\
Kinetic SZ   & 0.85 & 0.86 \\
Thermal SZ   & 3.90 & 4.40 \\
Dust         & 1.60 & 3.80 \\
Free-Free    & 0.30 & 0.38 \\
Synchrotron  & 0.05 & 0.06 \\
\hline
\end{tabular}
\label{tab1}
\end{center}
\end{table}

As an intermediate case, reconstructions were also performed assuming
that only the diagonal elements of the noise covariance matrix were
non-zero. This therefore includes information about the power spectra
of the instrumental noise and 
point source maps at each observing frequency, but ignores any
cross-correlation of the point source between frequencies. In this
case, the quality of the reconstructed maps is only slightly worse
and, in particular, the rms of the residuals in the CMB reconstruction
increases marginally to 8.3 $\mu$K.

In Fig.~\ref{ps1} we plot the power spectra of the reconstructed maps
(faint lines) and compare them with the power spectra of the true
input maps (bold lines) as shown in Fig.~\ref{inputs}. 
We see from Fig.~\ref{ps1} that,
in spite of the contamination due to point sources, the power spectrum
of the MEM reconstruction of the CMB component (panel a) is accurate
up to $\ell \approx 2000$, at which point the reconstruction begins to
underestimate the true power spectrum.  For the kinetic SZ effect
(panel b) the power spectrum of the reconstruction underestimates the
true spectrum at all multipoles, but for the thermal SZ effect (panel
c) the reconstructed spectrum follows the true one up to $\ell \approx
$1000.  The most accurately reconstructed power spectrum is that of
the Galactic dust emission (panel d), which closely follows the true
spectrum up to $\ell \approx 3000$. Finally, the power spectra of the
free-free and synchrotron reconstructions (panels e and f) slightly
underestimate the true spectra across all multipoles.  
It was further
found that assuming a diagonal noise covariance matrix did not
appreciably alter the results. 
As discussed in Paper I, the errors on
the reconstructed power spectra may be easily calculated, and we 
find that in each case the 68 percent confidence limits on the
reconstructed power spectra encompass the
true power spectrum at all multipoles. 

\subsection{Reconstructions with no power spectrum information}

For our second case, we take the opposite view to that adopted
above and assume that almost no power spectrum information is
available. This corresponds to assuming a flat (white-noise) power
spectrum for each component out to the highest measured Fourier
mode. The levels of the flat power spectra are chosen so that the
total power in each component is approximately that observed in the
input maps in Fig.~\ref{inputs}.  Furthermore, the noise covariance
matrix contains only the diagonal contribution form the instrumental
noise, and no attempt is made to include the effects due to point
source contamination. As discussed in Paper I, 
when no power spectrum information is assumed,
the MEM algorithm is iterated by using the current reconstructions
to update the ICF matrix $\mathbfss{L}$
before embarking on the next iteration. The solution was found to
converge after 8 such iterations.

\begin{figure*}
\centerline{\epsfig{
file=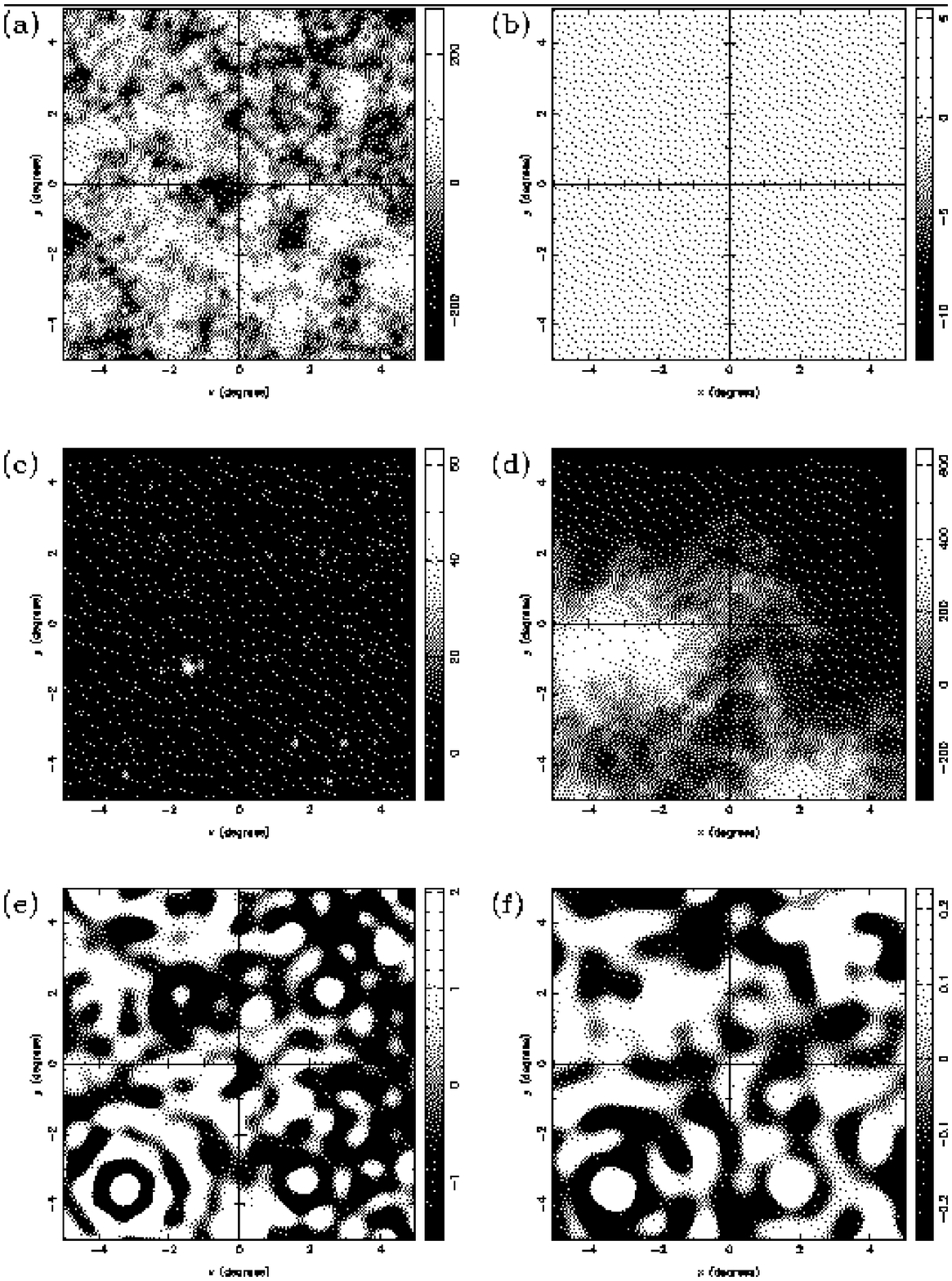,width=16cm}}
\caption{MEM reconstruction of the $10\times 10$ deg$^2$ maps of
the input components shown in Fig.~\ref{inputs}, using no power
spectrum information (see text). The components are:
(a) primary CMB fluctuations; (b) kinetic SZ effect (c) 
thermal SZ effect; (d) Galactic dust; (e) Galactic
free-free; (f) Galactic synchrotron emission. 
Each component is plotted at 300 GHz
and has been convolved with a Gaussian beam of FWHM equal to 4.5
arcmin. The map units are equivalent thermodynamic temperature in
$\mu$K.}
\label{maps2}
\end{figure*}

The resulting reconstructions are shown in Fig.~\ref{maps2}. 
\begin{figure*}
\centerline{\epsfig{
file=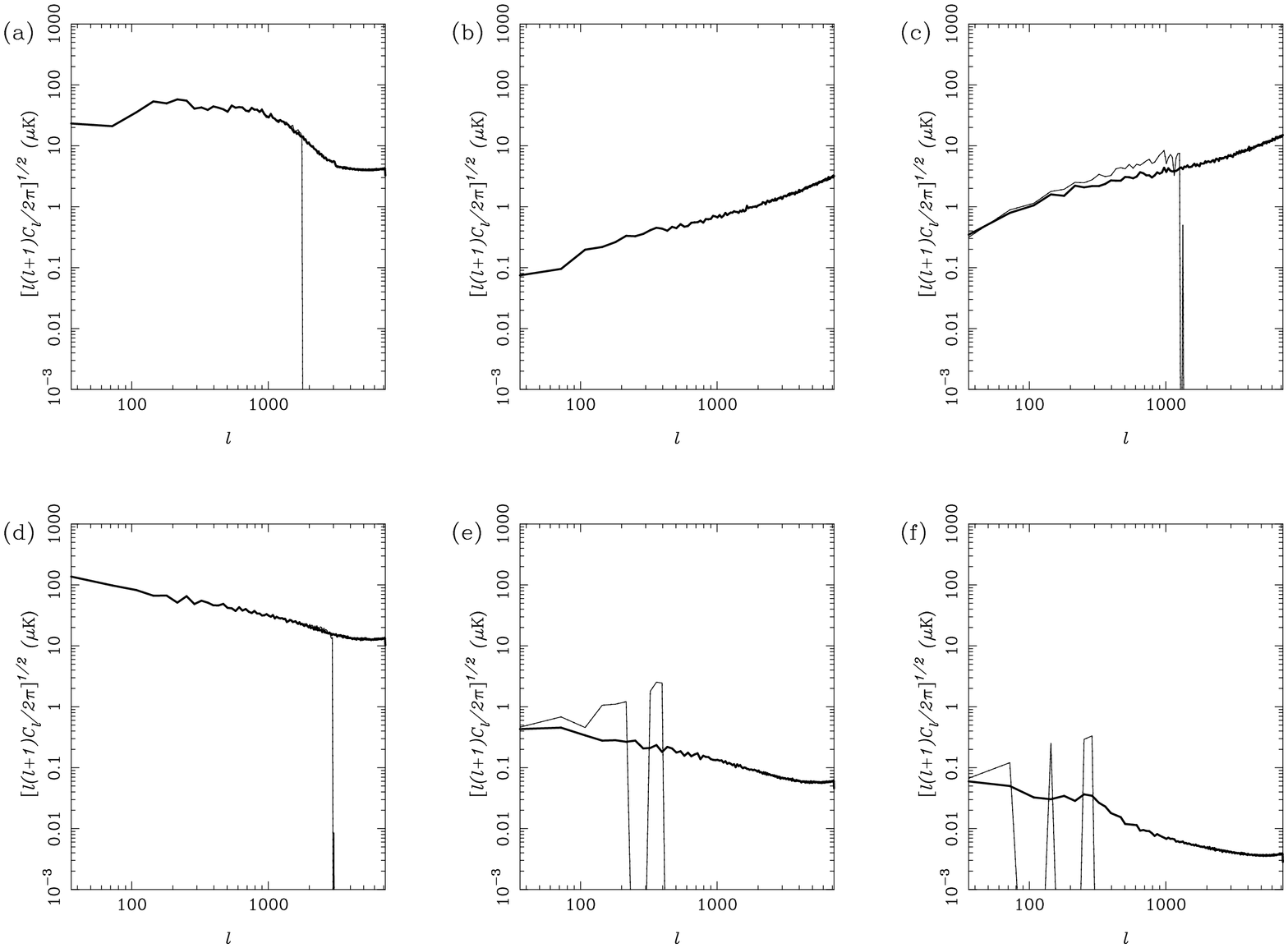,angle=-90,width=14.5cm}}
\caption{The power spectra of the input maps (bold line) compared to
to the power spectra of the maps reconstructed using MEM with no power
spectrum information (thin line). In panel (b) only the input power
spectrum is shown for the kinetic SZ effect, since the reconstructed
map in this case is everywhere zero.}
\label{ps2}
\end{figure*}
In this case, it is most encouraging that the CMB reconstruction is
still very accurate, even
for this very pessimistic scenario. Once again the reconstruction is,
at least by eye, almost indistinguishable from the true map shown in
Fig.~\ref{inputs}, except for the presence of a very bright point
source in the lower left-hand corner of the map. We also see that the
Galactic dust component is accurately recovered and contains very few
of the point sources present in the high-frequency channel maps shown
in Fig.~\ref{data2}. The thermal SZ effect has also been recovered,
but only in the brightest clusters, and the reconstrucion also
contains numerous misidentified point sources that appear as sharp
negative features. The reconstructions of the remaining three
components are very poor. In particular, the kinetic SZ reconstruction
has simply defaulted to zero in the absence of any useful data,
whereas the reconstructions of the Galactic free-free and synchrotron
emission are heavily contaminated with point sources and no obvious
features in the true input maps have been recovered.

In Table~\ref{tab2} we compare the rms residuals, $e_{\rm rms}$, per
4.5 arcmin FWHM Gaussian beam, of the
reconstructions shown in Fig.~\ref{maps2} with the corresponding
results obtained in Paper I in the absence of point sources.
\begin{table}
\begin{center}
\caption{The rms of the residuals per 4.5 arcmin FWHM Gaussian beam 
(in $\mu$K) of the MEM
reconstructions in the presence and absence of point sources, assuming
nopower spectrum information.}
\begin{tabular}{lcc}
\hline
Component & $e_{\rm rms}$ no sources & $e_{\rm rms}$ with sources \\
\hline
CMB          & 6.10 & 8.70 \\
Kinetic SZ   & 2.68 & 2.70 \\
Thermal SZ   & 4.35 & 5.80 \\
Dust         & 1.90 & 4.80 \\
Free-Free    & 0.44 & 1.60 \\
Synchrotron  & 0.07 & 0.23 \\
\hline
\label{tab2}
\end{tabular}
\end{center}
\end{table}
We see that the value of the $e_{\rm rms}$ has increased for all the 
components with respect to the case when point sources are not 
present in the simulations. Nevertheless, the relative errors for the
CMB, thermal SZ and Galactic dust components are still reasonably
small. In particular, we note that even in this pessimistic scenario,
the rms error on the CMB reconstruction is still only 8.7 $\mu$K.

Finally, in Fig.~\ref{ps2} we plot the power spectra of the
reconstructed maps and compare them with the input power spectra. 
For the CMB (panel a), we see that the reconstructed power spectrum again
follows the true spectrum very closely up to $\ell \approx 1800$, at
which point it drops directly to zero.  The dust power spectrum (panel
d) is also very accurately recovered up to $\ell \approx 2500$. For
the thermal SZ effect (panel c) follows the true spectrum up to $\ell
\approx 200$, but then begins gradually to overestimate the true
spectrum before dropping to zero at $\ell \approx 1200$. The
additional power in the reconstructed power spectrum is a result of
misidentifying point sources as thermal SZ effects from clusters. As
expected from the reconstructed maps shown in Fig.~\ref{maps2}, the
power spectra of the true kinetic SZ, free-free and synchrotron
emission are not recovered and the reconstructed free-free and
synchrotron power spectra oscillate widely about the true spectrum
before dropping to zero at $\ell \approx 300$.
We again find that the true power spectrum in each
case lies within the estimated one-sigma errors on the
reconstructed power spectrum at all multipoles.

We note that for the CMB and dust components the power spectra are
apparently recovered over a larger range of multipoles than when point
sources were not included in the analysis (see Paper I). The reason for
this becomes clear if we include full power spectrum information for
the six input components shown in Fig.~\ref{inputs}, but do not
include the point source contribution in the noise covariance matrix.
In this case, the reconstructed power spectra of the CMB and dust are
overestimated and possess a distict bump in the multipole range $\ell
\approx$ 1500--2500 and $\ell\approx$ 2500--3500 respectively.  This
feature is due to the contribution to high multipoles of the point
sources.  We must be careful when interpreting the power spectra of
the recovered maps at high multipoles, since these angular scales may
be contaminated by point source emission despite the fact that only a
few point sources are obvious in the reconstructed maps.

\section{Recovery of point source catalogues}

We have so far concerned ourselves primarily with the reconstruction
of the CMB and other foreground emission in the presence of point
sources. Nevertheless, an additional aim of the Planck Surveyor mission is to
compile point source catalogues over a wide range of frequencies.
In this section we therefore investigate how accurately we can recover
the input point source population used in the simulations.

It is clear that the generalised MEM approach discussed in Section
\ref{sepalg} has in no way been optimised to achieve the goal of recovering
point sources. Indeed, the mechanism for preventing the inclusion
of sources into the reconstructed CMB and diffuse foreground maps was
to regard the sources as an additional noise component. Nevertheless,
given the absence of point sources in the
reconstructions shown in Figs \ref{maps1} \& \ref{maps2}, 
it is clear that some information concerning these sources can be
recovered by comparing the reconstructions with the input data.

We use the following straightforward technique to obtain
information about the point source distribution at each Planck
Surveyor observing frequency. We first make simulated data at each
frequency using the MEM reconstructions as inputs (as opposed to true maps
shown in Fig.~\ref{inputs}). For each observing frequency the
MEM reconstruction of each physical components is projected in frequency
using its known spectral behaviour. The resulting maps are then added
together to produce the MEM reconstruction of the sky at this
frequency in the absence of point sources. This map is then convolved
with the appropriate Planck Surveyor observing beam to produce mock
data at this frequency. These
mock data are then subtracted from the true input data (which contain
the point source contribution). Since the MEM reconstructions
contain very few point sources, we thus hope to recover 
an approximate point source map
at each Planck Surveyor observing frequency.

\begin{figure}
\centerline{\epsfig{
file=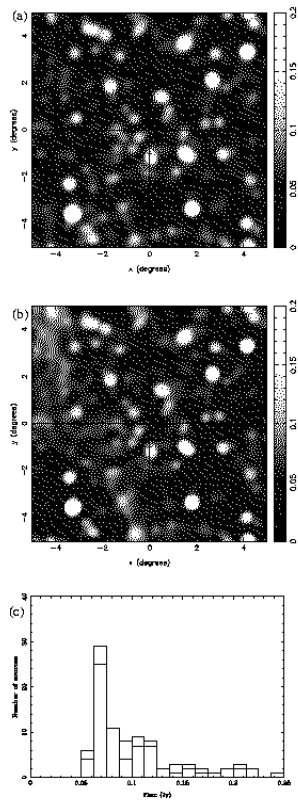,width=8cm}}
\caption{(a) The input point source map at 44 GHz convolved with the
appropriate Planck Surveyor observing beam. (b) The recovered point source 
map at 44 GHz. (c) A histogram of point source fluxes for the input
map (heavy line) and recovered map (faint line) obtained using
SExtractor; see text for details. The greyscales in (a) and (b)
are flux density in Jy}
\label{sourcerec1}
\end{figure}

\begin{figure}
\centerline{\epsfig{
file=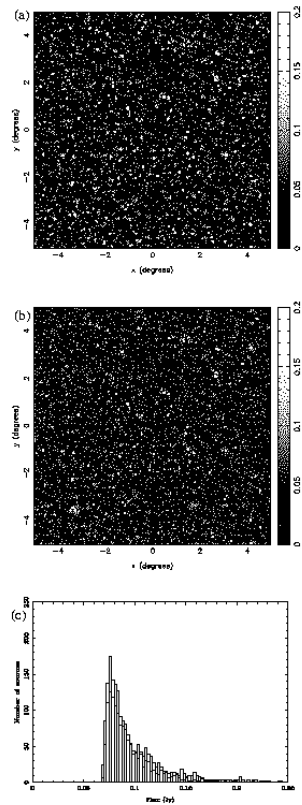,width=8cm}}
\caption{As for Fig.~\ref{sourcerec1} but for the 353 GHz
Planck Surveyor frequency channel.}
\label{sourcerec2}
\end{figure}

Using the MEM reconstructions shown in Fig~\ref{maps1}, which assume full
power spectrum information, we find that the recovered point source
maps are quite accurate. For the sake of brevity we display only
the reconstructions for
the 44 and 353 GHz channels, which are shown in
Figs \ref{sourcerec1} \& \ref{sourcerec2} respectively. 
These channels are chosen simply
because they lie near the centres of the frequency ranges of the LFI
and HFI respectively. We emphasise that these maps represent the point sources
as observed by the Planck Surveyor, i.e. convolved with the appropriate
observing beam.

In Fig.~\ref{sourcerec1}, we see that at 44 GHz the reconstructed point source
map in panel (b) is very similar to the input map in shown in panel
(a). Indeed all of the prominent point sources have been recovered at
approximately the correct flux. There is, however, some evidence
of residual diffuse emission in the top left-hand corner of
the recovered map. We quantify the point source recovery by
identifying and extracting point sources from the input and recovered
maps using the SExtractor algorithm (Bertin \& Arnouts
1996). This algorithm first fits an unresolved background to an image
and then identifies point sources superposed on this background.
Using the standard default settings, SExtractor fitted an unresolved
background with mean level 0.05 Jy per beam area to both the true and recovered
point source maps shown in Fig.~\ref{sourcerec1}(a) and (b). 
After subtracting this background, the resulting
rms fluctuation was found to be 0.01 Jy
for each map and SExtractor then identified sources down to this
level. Thus the detection flux limit in each case was 0.06 Jy.
For the input map, 93 sources were found as compared to 85 for
the recovered map. The resulting histograms of source counts for the input map
(heavy line) and the recovered map (faint line) are shown in 
Fig.~\ref{sourcerec1}(c) and are clearly very similar.

In Fig.~\ref{sourcerec2}, we show the results for the 353 GHz. Since the
FWHM of the observing beam at this frequency is only 4.5 arcmin, the
number of distinguishable point sources is much higher than at 44 GHz.
We note once more that the input and reconstructed point source maps are very
similar. Indeed, even very faint sources are accurately recovered.
SExtractor again fitted an unresolved background component with
mean value 0.05 Jy per beam area to both maps, resulting in
a residual rms fluctuation of 0.01 Jy in each case. Thus, as for the
44 GHz channel, the detection flux limit was 0.06 Jy. The algorithm
identified 1869 sources in the true map and
1676 sources in the recovered map. The histograms of the point source counts
are shown in Fig.~\ref{sourcerec2}(c) and are seen to agree reasonably
well. There is, however, some evidence that the recovered map has too
few sources in the range 0.10-0.12 Jy and too many in the range 0.06--0.08 Jy.
This is most likely caused by
some of the flux associated with the point
sources being incorrectly assigned to very low-level diffuse foreground
emission by the MEM separation algorithm.

The results at other observing frequencies are of similar quality to
those shown in Figs \ref{sourcerec1} \& \ref{sourcerec2}. 
We thus find that the input
point source catalogues at each Planck Surveyor frequency can be recovered to
reasonable accuracy using the above method. We re-emphasise that 
this has been achieved with no direct
subtraction of point sources from the data, but instead by the
application of the MEM separation algorithm in which the point sources
were considered as an extra `noise' contribution.

The recovered point source maps discussed above were created using
the MEM reconstructions obtained by assuming full power spectrum
information. Nevertheless, we find that the results are not
appreciably affected by using the MEM reconstructions obtained
assuming no power spectrum information shown in Fig.~\ref{maps2}.
At the lower LFI frequencies
the flux of the brightest sources is underestimated by about a factor
of two since some of flux has been misidentified by MEM as
synchrotron or free-free emission. In the HFI frequency range,
however, the dominant diffuse emission is due to the CMB and Galactic
dust, which are both accurately reconstructed, and the resulting
point source recovery is of a similar quality to that shown in
Fig.~\ref{sourcerec2}.

\section{Discussion and conclusions}

In this paper we study the effect of point sources on the
reconstruction of CMB anisotropies and foreground emission from
simulated satellite observations.  In particular, we apply a
generalised
form of the maximum-entropy method (MEM) 
developed in Hobson et al. (1998) to analyse simulated
observations by the Planck Surveyor satellite of a $10 \times
10$ deg$^2$ patch of sky. 

We do not attempt to remove 
contaminated pixel from the data, but instead
introduce information about the power spectrum of the point sources in
each frequency channel and the correlations between frequencies.
The resulting reconstructions show that, even
in the presence of contaminating point sources, the CMB emission
can still be recovered with an accuracy of about 8 $\mu$K, even for the
pessimistic scenario in which we assume no prior information
about the covariance structure of the sky emission. It is possible
that this accuracy could be improved by a suitable scheme for removing
point sources from the data before the MEM component separation
algorithm is applied. Moreover, given some prior knowledge of the power spectra
of the CMB and foregrounds, not only can the accuracy of the CMB
reconstruction be improved, but it is possible to obtain accurate 
reconstructions of the thermal SZ effect from clusters and the
Galactic dust, free-free and synchrotron emission. The kinetic SZ
effect may also be recovered in clusters that have a large thermal
effect.

As a final point, we find that by using the MEM reconstructions of the
CMB and foreground components it is possible to
subtract simulated mock data from the true data in order to
recover the point source contribution at each of the Planck Surveyor
observing frequencies. The recovered point source maps can then be
analysed using, for example, the SExtractor algorithm in order to
compile point source catalogues.

\section*{Acknowledgements}

JLS and RBB acknowledge financial support from the Spanish DGES,
project PB95-1132-C02-02, CICYT, project ESP96-2798-E and from
Comisi\'on Mixta Caja Cantabria-Universidad de Cantabria. RBB
acknowledges a Spanish M.E.C. Ph.D. scholarship and expresses her
gratitude to the Mullard Radio Astromomy Observatory for its
hospitality during her stay there.  LT acknowledges partial financial
support from the Spanish DGES, project PB95-1132-C02-02 and by the
Agenzia Spaziale Italiana (ASI).  LT also benefited from many years of
collaboration on foregrounds with L.~Danese, G.~De Zotti,
A.~Franceschini and C.~Burigana. FRB thanks R.~Gispert for permission
to use some of their unpublished results. We also thank E.~Mart\'\i
nez-Gonz\'alez, L.~Cay\'on, and F.~Arg\"ueso-G\'omez for their useful
comments.

\bsp  
\label{lastpage}

\begin{thebibliography}{99}
\bibitem[\protect\citename{Bersanelli et al. }1996]{bersanelli96}
Bersanelli M. et al., 1996, Report on Phase A Study for COBRAS/SAMBA, 
European Space Agency.
\bibitem[\protect\citename{Bertin \& Arnouts }1996]{bertin96}
Bertin E., Arnouts S., 1996, A\&AS, 117, 393
\bibitem[\protect\citename{Bouchet et al. }1997]{bouchet97}
Bouchet F.R., Gispert R., Boulanger F., Puget J.L., 1997, 
in Bouchet F.R., Gispert R., Guideroni B., Tran Thanh Van J.,eds, 
Proc. 36th Moriond Astrophysics Meeting, Microwave Anisotropies. 
Editions Fronti\`{e}re, Gif-sur-Yvette, p.~481
\bibitem[\protect\citename{Burigana et al. }1987]{burigana}
Burigana C., Danese L., De Zotti G., Franceschini A., Mazzei P.,
Toffolatti L., 1997, MNRAS, 287, L17
\bibitem[\protect\citename{Danese et al. }1987]{danese}
Danese L., De Zotti G., Francheschini A., Toffolatti
L., 1987, ApJ, 318, L15
\bibitem[\protect\citename{Francheschini et al. }1994]{francheschini}
Francheschini A., Mazzei P., De Zotti G., Danese L., 1994, ApJ, 427, 140
\bibitem[\protect\citename{Gispert \& Bouchet }1997]{gispert97}
Gispert R., Bouchet F.R., 1997, 
in Bouchet F.R., Gispert R., Guideroni B., Tran Thanh Van J.,eds, 
Proc. 36th Moriond Astrophysics Meeting, Microwave Anisotropies. 
Editions Fronti\`{e}re, Gif-sur-Yvette, p.~503
\bibitem[\protect\citename{Paper I }]{hjlb} 
Hobson M.P., Jones A.W., Lasenby A.N., Bouchet,F. 1998
MNRAS, in press (Paper I)
\bibitem[\protect\citename{Press et al. }1994]{nr}
Press P. H., Teukolsky S. A., Vettering W. T., Flannery B. P., 
1994, Numerical Recipes. Cambridge Univ. Press, Cambridge
\bibitem[\protect\citename{Puget et al. }1998]{puget}
Puget J.-L., Abergel A., Bernard J.-P., Boulanger F., Burton W.B.,
D\'esert F.-X., Hartmann D., 1996, A\&A, 308, L5
\bibitem[\protect\citename{Tegmark \& de Oliveira-Costa }1998]{tegandco}
Tegmark M., de Oliveira-Costa A., 1998, ApJ, submitted (astro-ph/9802123)
\bibitem[\protect\citename{Toffolatti et al. }1998]{toffolatti}
Toffolatti L., Arg\"ueso G\'omez F., De Zotti G., Mazzei P., 
Francheschini A., Danese L., Burigana C., 1998, MNRAS., 297, 117
\end{thebibliography}
\end{document}